\documentclass[10pt,letterpaper]{article}
\usepackage[footskip=0.75in,marginparwidth=2in]{geometry}

\usepackage[utf8]{inputenc}
\usepackage[T1]{fontenc} 

\usepackage{amsmath,amssymb}

\usepackage{cite}

\usepackage{nameref,hyperref}

\usepackage{microtype}
\DisableLigatures[f]{encoding = *, family = * }

\usepackage{ragged2e}

\usepackage{changepage}

\usepackage[aboveskip=1pt,labelfont=bf,labelsep=period,singlelinecheck=off]{caption}

\makeatletter
\renewcommand{\@biblabel}[1]{\quad#1.}
\makeatother

\usepackage{lastpage,fancyhdr,graphicx}
\usepackage{epstopdf}
\fancyheadoffset[L]{2.25in}
\fancyfootoffset[L]{2.25in}

\usepackage{color}
\definecolor{Gray}{gray}{.25}

\usepackage{graphicx}
\usepackage{sidecap}
\usepackage{wrapfig}
\usepackage[pscoord]{eso-pic}
\usepackage[fulladjust]{marginnote}
\reversemarginpar

\begin{document}
\justifying
\vspace*{4pt}

\begin{flushleft}
{\Large
\textbf{Moiré-Engineered Ferroelectric Transistors for Nearly Trap-free, Low-Power and Non-Volatile 2D Electronics}
}
\newline
\\
Arup Singha\textsuperscript{1,*},
Shaili Sett\textsuperscript{1},
Kenji Watanabe\textsuperscript{2},
Takashi Taniguchi\textsuperscript{2},
Arindam Ghosh\textsuperscript{2},
Rahul Debnath\textsuperscript{1,3,*},
\bigskip

\textbf{1} Department of Physics, Indian Institute of Science, Bangalore 560012, India
\\
\textbf{2} Research Center for Electronic and Optical Materials, National Institute for Materials Science, 1-1 Namiki, Tsukuba 305-0044, Japan
\\
\textbf{3} Department of Physics, National Institute of Technology, Agartala, Tripura 799046, India
\\
\bigskip
* arupsingha@iisc.ac.in, rahul.phy@faculty.nita.ac.in

\end{flushleft}

\section*{Abstract}
Long-range moiré patterns in twisted WSe$_2$ enable a built-in, moiré-length-scale ferroelectric polarisation that can be directly harnessed in electronic devices. Such a built-in ferroic landscape offers a compelling means to enable ultralow-voltage and non-volatile electronic functionality in two-dimensional materials; however, achieving stable polarization control without charge trapping has remained a persistent challenge. Here, we demonstrate a moiré-engineered ferroelectric field-effect transistor (FeFET) utilizing twisted WSe$_2$ bilayers that leverages atomically clean van der Waals interfaces to achieve efficient polarization–channel coupling and trap-suppressed, ultralow-voltage operation (subthreshold swing of $64~\mathrm{mV\,dec}^{-1}$). The device exhibits a stable non-volatile memory window of $0.10~\mathrm{V}$ and high mobility, exceeding the performance of previously reported two-dimensional FeFET and matching that of advanced silicon-based devices. In addition, capacitance--voltage spectroscopy, corroborated by self-consistent Landau--Ginzburg--Devonshire modelling, indicates ultrafast ferroelectric switching ($\sim0.5~\mu\mathrm{s}$). These results establish moiré-engineered ferroelectricity as a practical and scalable route toward ultraclean, low-power, and non-volatile 2D electronics, bridging atomistic lattice engineering with functional device architectures for next-generation memory and logic technologies.

\textbf{Keywords:} Moir\'{e} pattern, Transition metal dichalcogenides, ferroelectricity, Subthreshold swing

\section*{\textbf{Introduction}} 

The relentless demand for energy efficiency in modern electronics has pushed conventional MOSFETs toward their fundamental physical limits. Historically, performance discussions have centered on the subthreshold swing (SS)~\cite{chang2010practical,waldrop2016chips,waldrop2016more,franklin2015nanomaterials}, which is thermally constrained to 60~mV~dec$^{-1}$ at room temperature by the Boltzmann factor—an intrinsic ``Boltzmann tyranny'' \cite{zhou2017ferroelectric} that broadens the OFF--ON transition and necessitates higher operating voltages. Negative-capacitance concepts proposed in 2008\cite{salahuddin2008use} suggested that integrating ferroelectrics into the gate stack could amplify surface potential and yield sub-Boltzmann switching\cite{mcguire2017sustained}, motivating extensive exploration of ferroelectric oxides such as PZT and HfO$_2$-based dielectrics integrated with silicon FETs \cite{jain2014proposal,karda2015anti,dasgupta2015sub,khan2011ferroelectric,jo2016negative,ravikumar2024first,ravikumar2025first,li2023comparative,ham2020ferroelectric}. However, these approaches face persistent challenges related to interface traps, material integration, and dimensional scaling\cite{park2015ferroelectricity}. As devices continue to scale down, it has become evident that the dominant limitation is not the thermionic SS itself, but the trap-induced electrostatic inefficiencies that elevate power consumption and undermine reliable low-voltage switching.

This realization has shifted attention toward device platforms that naturally suppress interface traps while enabling ferroelectric functionality. Two-dimensional materials provide an atomically sharp, trap-clean channel with excellent electrostatic control, making them ideal hosts for ferroelectric field-effect transistors (FeFETs) in which polarization, rather than thermionic emission, modulates channel charge. Recent discoveries of intrinsic out-of-plane ferroelectricity in moiré superlattices\cite{jiang2025interplay,li2024sliding,kamaei2023ferroelectric,wu2024coexistence}—formed in twisted 2D bilayers such as WSe$_2$—offer a scalable and integration-friendly ferroic component fully compatible with van der Waals assembly\cite{jiang2025interplay,li2024sliding,kamaei2023ferroelectric,wu2024coexistence}. These ultraclean interfaces allow direct and efficient coupling of ferroelectric polarization to the channel, minimizing hysteresis from defect states and enabling stable, ultralow-voltage non-volatile operation. Such trap-suppressed ferroelectric gating provides memory windows as small as $\sim0.1~\mathrm{V}$, highlighting a promising route toward dense, low-voltage, and near-zero-standby-power memory and logic architectures.

To contextualise the significance of moiré ferroelectricity, it is essential to distinguish it from extrinsic ferroelectric gating using oxide or doped hafnia layers. Unlike externally deposited ferroelectrics, moiré-induced polarization arises intrinsically from periodic lattice reconstruction, leading to stable dipole ordering without chemical modification or epitaxial strain \cite{wang2022interfacial,weston2022interfacial}. This intrinsic nature provides an atomically clean route to integrate negative capacitance within van der Waals heterostructures, potentially lowering $V_{\mathrm{DD}}$ while preserving device scalability. In recent years, several research groups have made significant progress in addressing the integration challenges discussed above, thereby unlocking access to a range of emergent physical phenomena. Building on these advances, we have conducted a systematic investigation of the electronic and optoelectronic responses in 2D TMDC-based ferroelectric heterostructures, employing graphene as a sensing layer \cite{sett2024emergent,gill2025moire}.

Here, we demonstrate a two-dimensional moiré ferroelectric field-effect transistor (FeFET) based on a twisted WSe$_2$ bilayer, which harnesses intrinsic polarization to achieve robust and improved device performance. Our FeFET exhibits a near-ideal subthreshold swing of $\sim64$~mV~dec$^{-1}$ with minimal trap-induced broadening and a stable memory window of $0.10$~V. Self-consistent Landau-Ginzburg-Devonshire modelling quantifies the polarization-field relation, linking the observed hysteresis to fast ferroelectric switching dynamics in twisted WSe$_2$. Through sweep-rate-dependent hysteresis measurements, we probe the kinetics of ferroelectric domain switching and reveal that it is governed by a domain-wall-limited mechanism consistent with moiré-induced pinning. Capacitance spectra show a pronounced ferroelectric switching peak in $dC/dV_{\mathrm{g}}$, absent in control devices, and four distinct regimes---accumulation, depletion transition, ferroelectric switching, and inversion (deep depletion)---whose boundaries shift with frequency. Furthermore, we employ an equivalent-circuit model to rigorously analyse broadband capacitance--voltage spectroscopy data, enabling us to delineate the characteristic relaxation times of the ferroelectric layer ($\sim0.5~\mu\mathrm{s}$) from those of distinct interface trap populations. The quantum-capacitance slope reflects the gate-modulated density of states of MoS$_2$, while the low parallel conductance quantifies dielectric losses. This work establishes a quantitative framework for understanding the interplay between intrinsic moiré-engineered ferroelectricity and extrinsic interfacial dynamics, paving the way for low-power, steep-slope logic and nonvolatile memory \cite{mondal2019low} based on designer two-dimensional materials.

\textbf{Experimental Section}  

Atomically thin layers of MoS$_2$, WSe$_2$, graphite, and hBN were mechanically exfoliated onto SiO$_2$ (285~nm)/Si$^{++}$ substrates using the Scotch-tape method. Monolayers of MoS$_2$ and WSe$_2$ were identified by optical contrast and confirmed via Raman spectroscopy. To assemble the heterostructures, we employed a deterministic dry-transfer technique using a polycarbonate (PC) film on a polydimethylsiloxane (PDMS) hemispherical stamp, which enabled precise pick-up and release of individual flakes. The device stack (Fig.~1a) consisted of MoS$_2$ (channel) / hBN (dielectric, thickness $\approx 8~\mathrm{nm}$) / twisted bilayer (TBL) WSe$_2$ (ferroelectric) / graphite (back gate). The rotational alignment of the two WSe$_2$ layers was controlled by a high-precision rotational stage, achieving twist angles of $\sim$1$^\circ$, as established in Refs.~\cite{debnath2021simple,debnath2020evolution,debnath2022tuning}. Electrical contacts to the MoS$_2$ channel were defined using electron-beam lithography followed by Cr/Au (5~nm/50~nm) deposition, ensuring low-resistance ohmic contacts (Fig.~1a). Two nominally identical FeFET devices with comparable geometry were fabricated and characterized. For comparison, control devices were also fabricated: (i) MoS$_2$/hBN/graphite FETs without the TBL WSe$_2$ layer (see Supplementary Information), to isolate the ferroelectric contribution, and (ii) TBL WSe$_2$/hBN/graphite stacks for piezoresponse force microscopy (PFM) to confirm ferroelectric polarization switching. All electrical measurements were carried out under ambient conditions using a Keithley 4200 semiconductor parameter analyzer, and capacitance–voltage characteristics were measured over the 1~kHz--1~MHz frequency range with the CVU module.\\

\textbf{Ferroelectric Domain and Electrical Hysteresis}  

A strong piezoelectric response is observed in the twisted WSe$_2$ layer, as it is revealed by piezoresponse force microscopy in our previous work \cite{sett2024emergent} and highlights the existence of oppositely polarized domains in twisted TMDC (S6). A schematic in this regard is also described, the polar features as illustrated in Fig.~1b. The topography collected at high magnification clearly resolves alternate triangular domains of opposite phase, separated by non-polarized regions, confirming the presence of moiré-induced ferroelectric domains. While these static domain patterns establish the microscopic origin of polarization, their functional relevance becomes evident in the device transfer characteristics (Fig.~1c), where the reversible switching of these domains manifests as a well-defined but narrow hysteresis window in the $I_D$–$V_G$ curve with a positive threshold-voltage shift between forward and reverse sweeps ($\Delta V_{\mathrm{th}}>0$). Unlike conventional FeFETs \cite{he2016thermally,jeon2020hysteresis,yuan2024plane} where counterclockwise hysteresis arises from full ferroelectric switching in an n-type channel, the direction observed here reflects the unique field distribution \cite{si2019ferroelectric} across the MoS$_2$–hBN–twisted-WSe$_2$ stack. The finite thickness of the hBN dielectric screens the effective ferroelectric bias $V_{\mathrm{FE}} \approx V_g \cdot \frac{C_{\mathrm{FE}}}{C_{\mathrm{BN}} + C_{\mathrm{FE}}}$, enabling only partial polarization switching during sweeps. The moiré potential pins domains with a broad coercive-field distribution $p(E_c)$, such that the forward sweep ($-V_g \rightarrow +V_g$) reverses solely low-$E_c$ domains, while the reverse sweep ($+V_g \rightarrow -V_g$) benefits from residual polarization and inverted screening to activate higher-$E_c$ domains, yielding a net threshold shift
\[
\Delta V_{\mathrm{th}} \propto 
\int p(E_c) \cdot 
\frac{\tanh\!\left( V_{\mathrm{FE}} - E_c \right)}{\delta E_c} \, dE_c 
\Big|_{\mathrm{reverse - forward}},
\] 
where $p(E_c)$ is the coercive-field distribution arising from moiré disorder, and $\delta E_c$ reflects domain-wall mobility. Importantly, control devices without the twisted-WSe$_2$ layer, but MoS$_2$ on hBN alone, show negligible hysteresis, ruling out interface traps as the dominant mechanism (inset of Fig.~1c). The clean MoS$_2$/hBN interface, in contrast to MoS$_2$/SiO$_2$ devices where large trap-induced clockwise \cite{ali2021traps,kaushik2017reversible} loops are typically observed, further supports a ferroelectric origin. The observed hysteresis window corresponds to polarization switching between two stable remanent states. Using the equivalent series-capacitance model of the stack (schematic in Fig.~1d), the remanent polarization is extracted as $P_r = C_{\mathrm{eq}} \cdot \Delta V_{\mathrm{th}}/2$, yielding $P_r(3D) = 0.42~\mu\mathrm{C/cm^2}$, corresponding to $2.76\times10^{-12}~\mathrm{C/m}$. The coercive field is estimated as $E_c = \frac{\Delta V_{\mathrm{th}}/2}{t_{\mathrm{FE}}}$, giving $E_c \approx 0.09~\mathrm{V/nm}$. Those estimated values from measurement are in good agreement with previous reports \cite{wang2022interfacial,yasuda2021stacking}. These values, together with the narrow yet reproducible hysteresis window of $\sim0.10~\mathrm{V}$, combined with stable remanent polarization, demonstrate excellent non-volatile memory reliability in our moiré ferroelectric transistors.

\bigskip
\noindent

\textbf{Self-Consistent Polarization Dynamics} 

To quantitatively connect these nanoscale ferroelectric signatures with device operation, we analyzed the polarization–electric field ($P$–$E$) characteristics of the MoS$_2$/hBN/twisted-WSe$_2$ stack using Landau–Ginzburg–Devonshire (LGD) theory \cite{shen2020ferroelectric}. The ferroelectric free energy density is expressed as  $F(P) = \frac{\alpha P^2}{2} + \frac{\beta P^4}{4} + \frac{\gamma P^6}{6}$, where the Landau coefficients ($\alpha, \beta, \gamma$) are extracted by fitting to the experimental memory window and capacitance model. The field across the ferroelectric, $E_{\mathrm{FE}}$, is obtained self-consistently by combining LGD theory with electrostatics. The displacement field is given by  $D = \varepsilon_0 E_{\mathrm{hBN}} + P_{\mathrm{FE}} 
= \frac{\varepsilon_0 \varepsilon_{\mathrm{hBN}} V_{\mathrm{hBN}}}{t_{\mathrm{hBN}}}$, while the total gate voltage divides across the dielectric and ferroelectric as $V = V_{\mathrm{hBN}} + V_{\mathrm{FE}} 
= E_{\mathrm{hBN}} t_{\mathrm{hBN}} + E_{\mathrm{FE}} t_{\mathrm{FE}}$. Solving these relations yields
\[
E_{\mathrm{FE}}(P, V) = 
\frac{C_{\mathrm{hBN}} V - P}{\left( C_{\mathrm{hBN}} + C_{\mathrm{FE}} \right) t_{\mathrm{WSe_2}}},
\]
which shows that the effective field is screened as polarization increases. At equilibrium, the LGD restoring field balances this electrostatic field, giving a nonlinear equation in $P$ that we solve numerically using the Newton–Raphson method. Here, the ferroelectric free-energy density is modeled following the Landau--Ginzburg--Devonshire (LGD) formalism~\cite{hoffmann2018ferroelectric,majumdar2016revisiting}:
\begin{equation}
F_{P} = \alpha P^{2} + \beta P^{4} + \gamma P^{6},
\end{equation}
where $P$ is the polarization, and $\alpha$, $\beta$, and $\gamma$ are material-specific coefficients. For twisted WSe$_2$,
\begin{align}
\alpha &= 1.27\times10^{3}~\mathrm{J\,m\,C^{-2}}, \nonumber\\
\beta  &= 1.21\times10^{5}~\mathrm{J\,m^{5}\,C^{-4}}, \nonumber\\
\gamma &= 9.5\times10^{6}~\mathrm{J\,m^{9}\,C^{-6}}.
\end{align}

These parameters were extracted by fitting to the experimental memory window and capacitance spectra. The use of a sixth-order expansion ensures convergence near the coercive field and prevents unphysical polarization divergence under high bias. Absorbing those extracted parameters into the equation, we simulate the $P$–$E$ curve reflected in Fig.~1e, which highlights the coercive field of $0.09 \,\text{V}/\text{nm}$ and remanent polarization of $0.42 \,\mu\text{C}/\text{cm}^2$. In addition, the simulated polarization--electric field ($P$--$E$) curve exhibits a slight deviation from the ideal S-shaped hysteresis typically observed in ferroelectric materials. 
This deviation reflects the presence of a relatively shallow double-well potential landscape (as illustrated in Fig.~1f), which limits the accessibility of the negative-capacitance regime in our device structure due to capacitance mismatch.  \\

\textbf{Ferroelectric Control of Subthreshold Swing}

Building on the hysteresis behavior, the impact of moiré-engineered ferroelectricity on the switching characteristics is revealed through subthreshold swing (SS) measurements. Figure 2a–c present the transfer characteristics, revealing a hysteresis that remains essentially independent of $V_{\mathrm{DS}}$, accompanied by a progressive improvement in subthreshold swing (SS) from $\sim 75$ to $\sim 64$ mV dec$^{-1}$ with increasing $V_{\mathrm{DS}}$.
Figure 2d shows the point-by-point subthreshold swing (SS) as a function of drain current, revealing an extended current range over which the SS approaches the fundamental 60 mV dec$^{-1}$ limit. Forward and backward sweeps yield nearly identical SS values, with only a marginal advantage for the backward sweep. In ferroelectric transistors where traps dominate, backward sweeps typically show a pronounced SS improvement due to trap neutralization followed by enhanced ferroelectric polarization. The absence of such a signature here indicates that trap states are minimally involved in the hysteresis mechanism. Instead, the bias-independent hysteresis suggests that the twisted-WSe$_2$ ferroelectric layer maintains a stable polarization over the measurement cycle, with negligible screening by trapped charges in the gate stack heterostructure.

Further confirmation arises from our previous work Ref.~\cite{sett2024emergent} using dual-gate measurements on a graphene channel, where the top-gate dielectric (hBN) exhibits no detectable hysteresis, while the back-gate incorporating twisted WSe$_2$ shows a clear hysteresis loop. This contrast unambiguously attributes the hysteresis to the ferroelectric polarization of the twisted WSe$_2$, rather than trap-assisted effects. Compared to a reference MoS$_2$/hBN device without the twisted-WSe$_2$ interlayer, which shows SS values exceeding $130$~mV\,dec$^{-1}$, the present device approaches the thermionic limit of $60$~mV\,dec$^{-1}$ across much of the subthreshold regime. The SS variation with drain current remains close to ideal, indicating strong gate control and minimal trap-induced broadening of the subthreshold slope. This highlights the role of the twisted-WSe$_2$ layer in enabling stable, low-trap, nonvolatile electrostatic modulation of the MoS$_2$ channel. To assess device uniformity, Fig.~S3 and S4 (Supplementary Information) presents transfer characteristics for three FeFETs fabricated under identical conditions, exhibiting subthreshold swings of $63$–$66~\mathrm{mV~dec}^{-1}$ and memory windows between $0.09$–$0.11~\mathrm{V}$. This consistency confirms the reproducibility of the fabrication method.\\

\textbf{Rate-Dependent Switching Kinetics}

To further probe the origin of hysteresis, we performed sweep-rate–dependent transfer measurements of the FeFET. Figure 3a shows that the hysteresis width ($H$) systematically decreases with increasing sweep rate, consistent with a domain-wall–limited ferroelectric switching process rather than trap-controlled dynamics. The dependence of $H$ on sweep rate ($r$) is well captured by a stretched-exponential formalism \cite{jo2009nonlinear,yudin2020modeling}, $H(r) = H_{\mathrm{sat}} \, e^{-(r/r_c)^n}$, where $H_{\mathrm{sat}}$ is the saturation hysteresis at vanishing sweep rate, $r_c$ the critical rate at which the switching fraction is reduced by $1/e$, and $n$ characterizes the breadth of the domain-wall velocity distribution. Experimental data (Fig.~3b, symbols) are well reproduced by the fitted curve, yielding $H_{\mathrm{sat}} = 5.00$ V, $r_c = 0.205$ V h$^{-1}$, and $n = 0.285$. The large $H_{\mathrm{sat}}$ indicates nearly complete polarization reversal under quasi-static biasing, whereas the small $r_c$ suggests moderate domain-wall mobility with switching suppressed even at relatively slow ramp rates. The broad exponent ($ n\ll 1$) implies a wide distribution of domain-wall velocities, consistent with pinning from moiré-induced strain fields or interfacial disorder. Together, these results confirm that the observed hysteresis originates predominantly from intrinsic ferroelectric switching, with charge trapping playing only a negligible role.

\textbf{Capacitance Spectroscopy of 2D FET}

To elucidate the underlying charge modulation mechanisms, capacitance--voltage ($C$-- $V_g$) measurements were performed on MoS$_2$/hBN devices both with and without the ferroelectric twisted WSe$_2$ layer in the gate stack. The FeFET configuration (MoS$_2$/hBN/twisted-WSe$_2$) exhibited a pronounced enhancement in capacitance (Figure 4a) compared to the control MOSFET (MoS$_2$/hBN). This enhancement is consistent with the additional charge contribution from polarization switching in the ferroelectric layer. The derivative $dC/dV_g$ serves as a sensitive probe for this switching \cite{jo2009nonlinear,hoffmann2018ferroelectric}, revealing a single, sharp, and symmetric peak in the FeFET curve (Figure 4b) with a full width at half maximum (FWHM) of $0.80~\mathrm{V}$ and a peak amplitude of $2.3~\mathrm{pF/V}$. This signature is a direct fingerprint of domain-wall-mediated ferroelectric switching. In contrast, the control MOSFET displayed a much broader and less intense peak (FWHM = $1.14~\mathrm{V}$, amplitude = $0.95~\mathrm{pF/V}$), characteristic of non-cooperative, trap-mediated electrostatic modulation. This is consistent with a conventional MOS capacitor structure, where the capacitance transitions from accumulation to depletion and eventually saturates at high negative $V_g$, representing the deep depletion or inversion capacitance.

Further insights are gained from the device's frequency-dependent response. At low frequencies ($30~\mathrm{kHz}$), the $C$--$V_g$ curve exhibits four characteristic regimes (Figure 4c) : (i) a high-capacitance accumulation region ($3$ to $5~\mathrm{V}$), (ii) a depletion transition ($1$ to $3~\mathrm{V}$), (iii) a ferroelectric switching window ($\approx -0.2~\mathrm{V}$ to $1~\mathrm{V}$) where polarization reversal in the twisted WSe$_2$ layer induces negative capacitance effects, followed by (iv) a post-switching decrease in capacitance from $-2.8~\mathrm{V}$ to $-0.2~\mathrm{V}$, signifying completion of ferroelectric switching as the depletion region expands into the semiconductor channel. At high negative gate voltages ($< -2.8~\mathrm{V}$), the device is either in strong inversion (formation of a hole inversion layer in the MoS$_2$/WSe$_2$ channel) or deep depletion. The total capacitance of the FeFET significantly decreases from a maximum of $5.25~\mathrm{pF}$ at $1~\mathrm{kHz}$ to $1.8~\mathrm{pF}$ at $1~\mathrm{MHz}$. This frequency suppression is a hallmark of relatively slow-response phenomena, namely ferroelectric polarization reversal and the contribution of interface trap states. Simultaneously, the ferroelectric switching peak in $dC/dV_g$ vanishes at high frequencies, indicating kinetic limitations in domain-wall motion under rapid gate modulation (Fig.~4d). Above several hundred kilohertz, the device response converges to the dielectric-only limit, confirming the ferroelectric origin of the low-frequency enhancement. From the minimum capacitance in the subthreshold regime, we infer that the capacitance floor is set by the depletion capacitance $C_d$, which remains finite even at low carrier densities. The absence of a vanishing capacitance rules out a dominant quantum capacitance $C_q$ contribution in this regime, as such an effect would appear in series with $C_d$ and further reduce the total capacitance. To gain a deeper understanding of the total measured capacitance, \( C_{\text{total}}(\omega, V_g) \), in a gated two-dimensional semiconductor device is modeled as a series combination of the oxide capacitance of the top-gate \( C_{\text{TG}} \), the quantum capacitance \( C_Q \), and the frequency-dependent interface trap capacitance \( C_{it}(\omega) \):
\begin{equation}
\frac{1}{C_{\text{total}}} = \frac{1}{C_{\text{TG}}} + \frac{1}{\left(C_Q + C_{it}(\omega)\right)}.
\label{eq:Ctotal}
\end{equation}

The interface-trap contribution is represented by two characteristic branches with distinct time constants \( \tau_A \) and \( \tau_B \), expressed as
\begin{equation}
C_{it}(\omega) = e^2 \left[
D_{it,A} \frac{\tan^{-1}(\omega \tau_A)}{\omega \tau_A}
+ D_{it,B} \frac{\tan^{-1}(\omega \tau_B)}{\omega \tau_B}
\right],
\label{eq:Cit}
\end{equation}
where \( D_{it,A} \) and \( D_{it,B} \) are the trap densities (in eV\(^{-1}\)cm\(^{-2}\)), \( e \) is the elementary charge, and \( \omega = 2\pi f \) is the angular frequency. The extracted quantum capacitance \(C_Q\) and the characteristic trap time \(\tau_{\mathrm{it}}\) obtained from the constrained global fit are plotted as a function of \(V_{\mathrm{TG}}-V_{\mathrm{TH}}\) in Fig.~4(f).

\section*{Trap Density analysis }
The interface trap density ($D_{\mathrm{it}}$) near the semiconductor--dielectric interface was first estimated from the subthreshold swing (SS) \cite{jeon2020hysteresis} characteristics using
\begin{equation}
SS = \left( 1 + \frac{q D_{\mathrm{it}}}{C_{\mathrm{ox}}} \right) \frac{kT}{q} \ln(10),
\end{equation}
which reflects the trap density at the Fermi level (Fig.~4(g)) when the device is near turn-on.
For the twisted--WSe$_2$ FeFET, the extracted $SS = 64~\mathrm{mV/dec}$ corresponds to 
$D_{\mathrm{it}} = 2.8\times10^{11}~\mathrm{cm^{-2}\,eV^{-1}}$, 
whereas the control device without the ferroelectric layer ($SS = 130~\mathrm{mV/dec}$) yields 
$D_{\mathrm{it}} = 3.0\times10^{12}~\mathrm{cm^{-2}\,eV^{-1}}$. 
This one-order-of-magnitude reduction in interface trap density indicates an ultraclean interface in the twisted--WSe$_2$ gated structure, consistent with the suppression of interfacial disorder and charge scattering. To further validate the trap response, high--low frequency \cite{yu2023simultaneously,zhao2018evaluation} $C$--$V$ spectroscopy was performed. 
Using the standard formulation \cite{yu2023simultaneously}
\begin{equation}
C_{\mathrm{it}} =
\left( 
  \frac{1}{C_{\mathrm{LF}}} - \frac{1}{C_{\mathrm{ox}}}
\right)^{-1}
-
\left( 
  \frac{1}{C_{\mathrm{HF}}} - \frac{1}{C_{\mathrm{ox}}}
\right)^{-1},
\end{equation}
the extracted $D_{\mathrm{it}}$ values (Fig.~4e) were slightly higher than those obtained from the SS analysis, consistent with additional slow interface states that respond at lower frequencies. 
Frequency-dependent capacitance measurements at various gate voltages revealed a linear dispersion with $\log(f)$ beyond the ferroelectric switching regime, 
indicating that the apparent frequency response in the accumulation region cannot be attributed to border traps. 
Instead, the observed dispersion originates from ferroelectric polarization dynamics. 
This ferroelectric contribution was quantitatively modeled using a constrained Cole--Cole magnitude formalism to extract the characteristic relaxation time ($\tau_{\mathrm{FE}}$) and dispersion parameter ($\alpha_{\mathrm{FE}}$) (see Supplementary Information).

\section*{Equivalent Circuit Analysis }
The equivalent circuit employed to analyze the small-signal response of the top-gated monolayer MoS$_2$ field-effect transistors (FETs) is illustrated in Figure~5a along with the device schematic in Figure~5b. 
The model comprehensively captures both intrinsic and extrinsic contributions to the total measured capacitance, enabling a quantitative interpretation of parasitic effects and channel dynamics. 
The circuit consists of the intrinsic gate-stack capacitances—namely, the oxide capacitance ($C_{\mathrm{ox}}$) in series with the quantum capacitance ($C_Q$)—in parallel with the interface trap branch characterized by the capacitance ($C_{\mathrm{it}}$) and resistance ($R_{\mathrm{it}}$), which represent carrier capture and emission processes at the dielectric/semiconductor interface. 
The total device resistance is separated into the gate-modulated channel resistance ($R_{\mathrm{ch}}$) under the gate and the bias-independent access resistance ($R_{\mathrm{c}}$), which includes both the MoS$_2$ sheet resistance in the ungated region and the metal–semiconductor contact resistance. 

To account for the finite $R_{\mathrm{c}}$ that limits the measured capacitance at high frequencies, the experimental response was modelled using an equivalent parallel configuration \cite{fang2018accumulation} given by
\begin{equation}
    C_{\mathrm{para}} = \frac{C_{\mathrm{ideal}}}{1 + (\omega R_{\mathrm{c}} C_{\mathrm{ideal}})^2},
\end{equation}
where $C_{\mathrm{ideal}}$ represents the ideal gate capacitance in the absence of interface traps and parasitic effects, and $\omega = 2\pi f$ is the angular frequency. 
This formulation captures the attenuation of the apparent capacitance in the accumulation regime as $R_{\mathrm{c}}$ increases. The simulation neglects other resistive components such as $R_{\mathrm{ch}}$, which are effectively shunted under strong accumulation. The extracted $C_{\mathrm{para}}$ (Figure~5c) asymptotically approaches $C_{\mathrm{ox}}$ at 1~MHz when $R_{\mathrm{c}} \approx 10^6~\Omega$, validating the consistency of the equivalent circuit model and confirming that the measured suppression of the capacitance in the accumulation regime primarily originates from the series resistance in the ungated MoS$_2$ access regions.

\section*{Quantum capacitance analysis }
Based on the ideal equivalent circuit model\cite{chen2025quantum} in the high-frequency limit, where the interface trap capacitance ($C_{\mathrm{it}}$) is negligible, the theoretical correlation between the channel potential ($V_{\mathrm{CH}}$) and the applied gate voltage ($V_{\mathrm{g}}$) can be expressed as
\[
V_{\mathrm{g}} = V_{\mathrm{g,mid\text{-}gap}} + \int_0^{V_{\mathrm{CH}}} \frac{C_{\mathrm{Q}} + C_{\mathrm{TG}}}{C_{\mathrm{TG}}}\, dV_{\mathrm{CH}},
\]

where $V_{\mathrm{g,mid\text{-}gap}}$ is a fitting parameter corresponding to the gate voltage at which the Fermi level lies at the mid-gap ($E_{\mathrm{F}} = 0~\mathrm{eV}$). This term effectively accounts for the intrinsic $n$-type doping present in monolayer MoS$_2$. By combining this relation with the theoretically derived expression of the quantum capacitance ($C_{\mathrm{Q}}$), the variation of $C_{\mathrm{Q}}$ as a function of the gate voltage ($V_{\mathrm{g}}$)  (Figure~5c inset) can be quantitatively determined and then compared with a microscopic two-dimensional Fermi-occupation model. Experimentally, the oxide capacitance \(C_{\mathrm{ox}}\) was estimated from the accumulation plateau, and the quantum capacitance was recovered using the series relation
\[
\frac{1}{C_{\mathrm{tot}}}=\frac{1}{C_{\mathrm{ox}}}+\frac{1}{C_Q}\,.
\]
The theoretical curve was computed for a 2-D conduction band with a constant band-edge density of states
\[
g_{2\mathrm{D}}=\frac{g_s g_v m^*}{2\pi\hbar^2},
\]
using \(m^*=0.6\,m_0\), \(g_s=g_v=2\) and a conduction-band edge \(E_c=0.95\ \mathrm{eV}\) referenced to the midgap \(E_F=0\). The sheet density and quantum capacitance were evaluated from Fermi statistics as
\begin{equation}
n(\mu) = g_{2\mathrm{D}} k_{\mathrm{B}} T 
\ln\!\left[ 1 + \exp\!\left( \frac{\mu - E_{\mathrm{c}}}{k_{\mathrm{B}} T} \right) \right],
\label{eq:n_mu}
\end{equation}

\begin{equation}
C_{Q} = e^{2} \frac{\partial n}{\partial \mu} 
= \frac{e^{2} g_{2\mathrm{D}}}{1 + \exp\!\left( \frac{E_{\mathrm{c}} - \mu}{k_{\mathrm{B}} T} \right)}.
\label{eq:CQ}
\end{equation}

This combined procedure provides a direct, physically transparent test of the DOS-derived quantum capacitance and provides information on interface traps, incomplete accumulation, or misestimated oxide coupling — immediately apparent, offering useful diagnostics for 2-D FeFET/TFET device modelling.

To quantitatively interpret the capacitance--voltage characteristics, we modeled the system using an equivalent circuit \cite{hoffmann2018ferroelectric,majumdar2016revisiting} that accounts for ferroelectric switching (Figure 4d), interface traps, quantum capacitance, and conductance losses:
\begin{equation}
\boxed{
\frac{1}{C_{\mathrm{total}}(V,\omega)} 
= 
\frac{1}{C_{g}(V,\omega)} 
+ 
\frac{1}{
C_{Q}(V)
+ 
\displaystyle\sum_{i=1}^{2}
\frac{C_{t,i}(V)}{1 + (\omega \tau_{i})^{2}}
}
}
\label{eq:Ctotal_model}
\end{equation}

\vspace{3mm}

\begin{equation}
C_{g}(V,\omega)
= 
\left(
\frac{1}{C_{\mathrm{ox}}}
+
\frac{1}{C_{\mathrm{FE}}(V,\omega)}
\right)^{-1},
\label{eq:Cg_model}
\end{equation}

\vspace{3mm}

\begin{equation}
C_{\mathrm{FE}}(V,\omega)
= 
\frac{C_{\mathrm{FE},0}(V)}{1 + (\omega \tau_{\mathrm{FE}})^{2}}.
\label{eq:CFE_model}
\end{equation}

The equivalent-circuit model \cite{fang2018band} (Figure~5d) accounts for ferroelectric polarization ($C_{\mathrm{FE}}$), interface traps ($C_{t1}$, $C_{t2}$), and quantum capacitance ($C_{q}$), enabling a self-consistent interpretation of the $C$–$V$ response~\cite{hoffmann2018ferroelectric}. 
The extracted relaxation times ($\tau_{\mathrm{FE}} = 0.5~\mu\mathrm{s}$, $\tau_{1} = 6.0\times10^{-3}~\mathrm{s}$, $\tau_{2} = 1.0\times10^{-7}~\mathrm{s}$) separate intrinsic ferroelectric switching from interface-mediated processes, emphasizing that the dominant kinetics originate from domain-wall motion within the moiré potential landscape. The fitting uncertainty of the extracted relaxation times was within $\pm10\%$ across multiple frequency sweeps, and repeated CV measurements yielded consistent spectra, confirming that the observed frequency dependence originates from intrinsic ferroelectric dynamics rather than measurement artifacts.

\section*{Benchmarking Device performance}

The overall device performance of the twisted--WSe$_2$ FeFET was benchmarked in terms of both switching steepness and polarization strength (Fig.~5). 
The device exhibits a subthreshold swing of $64~\mathrm{mV\,dec^{-1}}$, approaching the sub-Boltzmann limit, 
along with a remanent polarization of $P_{\mathrm{r}}\approx0.42~\mu\mathrm{C/cm^2}$ and a coercive field of $E_{\mathrm{c}}\approx0.09~\mathrm{V/nm}$. 
These values substantially outperform the control MoS$_2$/hBN transistor ($SS = 130~\mathrm{mV\,dec^{-1}}$) and compare favorably with state-of-the-art twisted or moiré ferroelectric transistors based on In$_2$Se$_3$, hBN, and MoS$_2$ heterostructures. The enhanced charge–polarization coupling and improved electrostatic control underscore the effectiveness of moiré engineering in achieving steep-slope and low-voltage operation in two-dimensional ferroelectric field-effect transistors.

A comparative analysis of reported moiré ferroelectric systems reveals that the polarization magnitude and coercive field exhibit strong dependence on interlayer coupling and twist angle. The relatively high remanent polarization observed in the twisted--WSe$_2$ device arises from enhanced interfacial dipole alignment and efficient domain-wall motion across the moiré potential landscape. Simultaneously, the moderate coercive field facilitates low-power switching without polarization fatigue. Together, these characteristics place the present device among the best-performing van der Waals ferroelectric transistors reported to date, bridging the gap between steep-slope logic and nonvolatile memory functionalities within a unified two-dimensional platform.

\section*{Conclusion}
Our findings reveal that moiré engineering in 2D heterostructures can deliver simultaneously steep-slope switching and robust non-volatile memory, directly addressing the energy-delay trade-off in future electronics. The switching dynamics—driven by domain-wall motion pinned within the moiré potential—combine rapid response ($\sim0.5~\mu\mathrm{s}$ relaxation) with long-term stability, a balance essential for both high-speed logic and reliable data retention. By disentangling ferroelectric and interfacial effects through capacitance–voltage spectroscopy, we establish a quantitative framework for rational FeFET design. Our findings open a route toward integrating moiré ferroelectrics with scalable CMOS-compatible architectures. Future work should explore replacing hBN with high-$\kappa$ dielectrics to enhance gate coupling, as well as exploring heterobilayer combinations (e.g., MoSe$_2$/WSe$_2$) for tunable polarization landscapes. Such advancements could enable sub-50~mV\,dec$^{-1}$ switching and multi-bit memory functionalities with reconfigurable neuromorphic circuits, positioning moiré-engineered FeFETs as key candidates for beyond-Boltzmann logic circuits.

\section*{Experimental method}
All electrical measurements were performed in an ultrahigh vacuum cryogenic probe station ($10^{-6}~\mathrm{mbar}$) using a Keithley 4200 semiconductor parameter analyzer with high-sensitivity triaxial cable to reduce noise and ultralow current measurements of 0.1 fA resolution. Capacitance-voltage (C–V) measurements spanned frequencies from 1~kHz to 1~MHz using the CVU module, with the device grounded at source and back gate held at a constant potential. To ensure reproducibility, each measurement was repeated at least three times, and device statistics were compiled across all runs. Supplementary Information includes raw transfer curves, control device measurements, and details of instrument calibration.

\section*{Supplementary Information}
Supplementary data include: (i) raw $I_D$–$V_G$ and $C$–$V_G$ characteristics for all FeFETs, (ii) parameter fitting workflow for the LGD and equivalent-circuit models, (iii) additional PFM confirmation of domain orientation, and (iv) details of the Self-Consistent Polarization Model and Capacitance spectroscopy.

\newpage
\section*{Author Contributions}
R.D. and A.S. designed all the experiments. A.S. and R.D. completed fabrication and characterization, A.S. and R.D. performed the measurements. R.D. and A.S. wrote the manuscript and analyzed the data, and performed device simulations. S.S. contributed to reviewing and editing the manuscript. A.G. supervised the overall project, edited, and revised the manuscript. All authors discussed the results and commented on the manuscript.

\section*{Code availability}
The computer code used to generate the results reported in this study is available from the corresponding author upon request.

\section*{Competing interests}
The authors declare no conflict of interest.

\section*{ACKNOWLEDGMENTS}
The authors acknowledge the usage of NNFC facilities at CeNSE, IISc, and funding from the Department of Science and Technology (DST), Govt. of India.  A.G. acknowledges the J.C. Bose Fellowship (Grant number SP/DSTO-18-2038) from Science and Engineering Research Board, DST and Nano Mission, DST, Govt. of India for financial support under Grant No.~DST/NM/TUE/QM-10/2019. S.S. would like to acknowledge MEITY, Govt. of India (Project Number: SP/MEIT-24-0007) for financial support.  K.W. and T.T. acknowledge support from the JSPS KAKENHI (Grant Numbers 21H05233 and 23H02052), the CREST (JPMJCR24A5), JST and World Premier International Research Center Initiative (WPI), MEXT, Japan.

\begin{figure*} 
\includegraphics[width=1\linewidth]{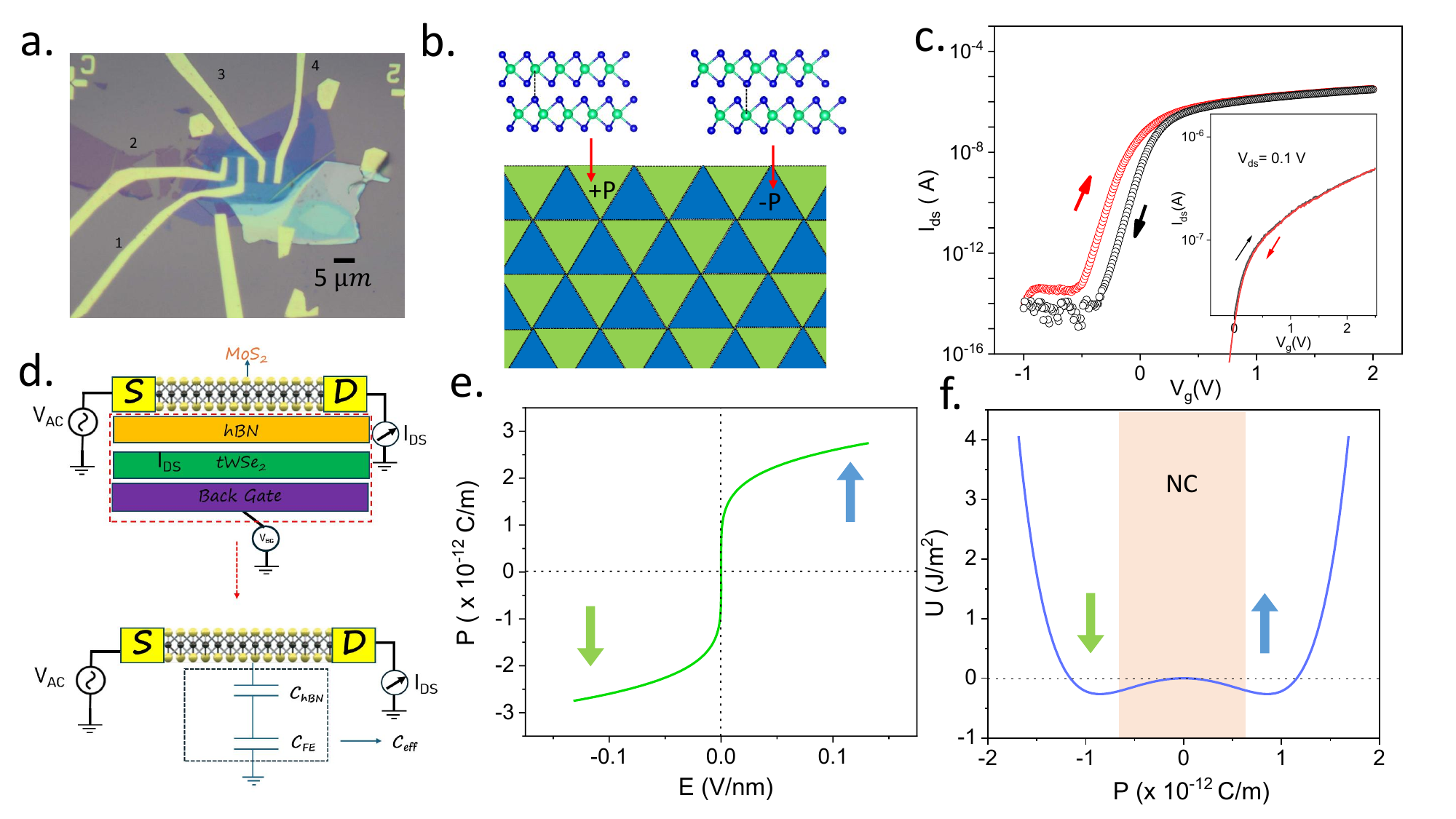} 
\caption{\textbf{Moiré ferroelectricity and device characteristics} 
(a) Optical micrograph of a representative MoS$_2$/hBN/twisted-WSe$_2$/graphite device. (b) Illustration of ferroelectric domains in twisted WSe$_2$. Green and blue colors represent up and down polarization, respectively. (c)Transfer characteristics ($I_{D}$–$V_{G}$) show a narrow clockwise hysteresis window ($\Delta V_{\mathrm{th}} \approx 0.10~\mathrm{V}$), while the control MoS$_2$/hBN/graphite device exhibits negligible hysteresis, excluding interface traps as the dominant mechanism (c, inset). (d) A schematic of the FeFET stack and the equivalent capacitance model is shown. (e) The calculated polarization–electric field ($P$–$E$) loop using the Landau–Khalatnikov formalism matches the extracted remanent polarization ($P_{r} \approx 0.42~\mu \mathrm{C\,cm^{-2}}$) and coercive field ($E_{c} \approx 0.09~\mathrm{V\,nm^{-1}}$), while the corresponding Landau free-energy profile (f) illustrates the characteristic double-well potential, confirming robust ferroelectric switching in twisted WSe$_2$.}
\label{fig:transport1}
\end{figure*}

\begin{figure*} 
\includegraphics[width=1\linewidth]{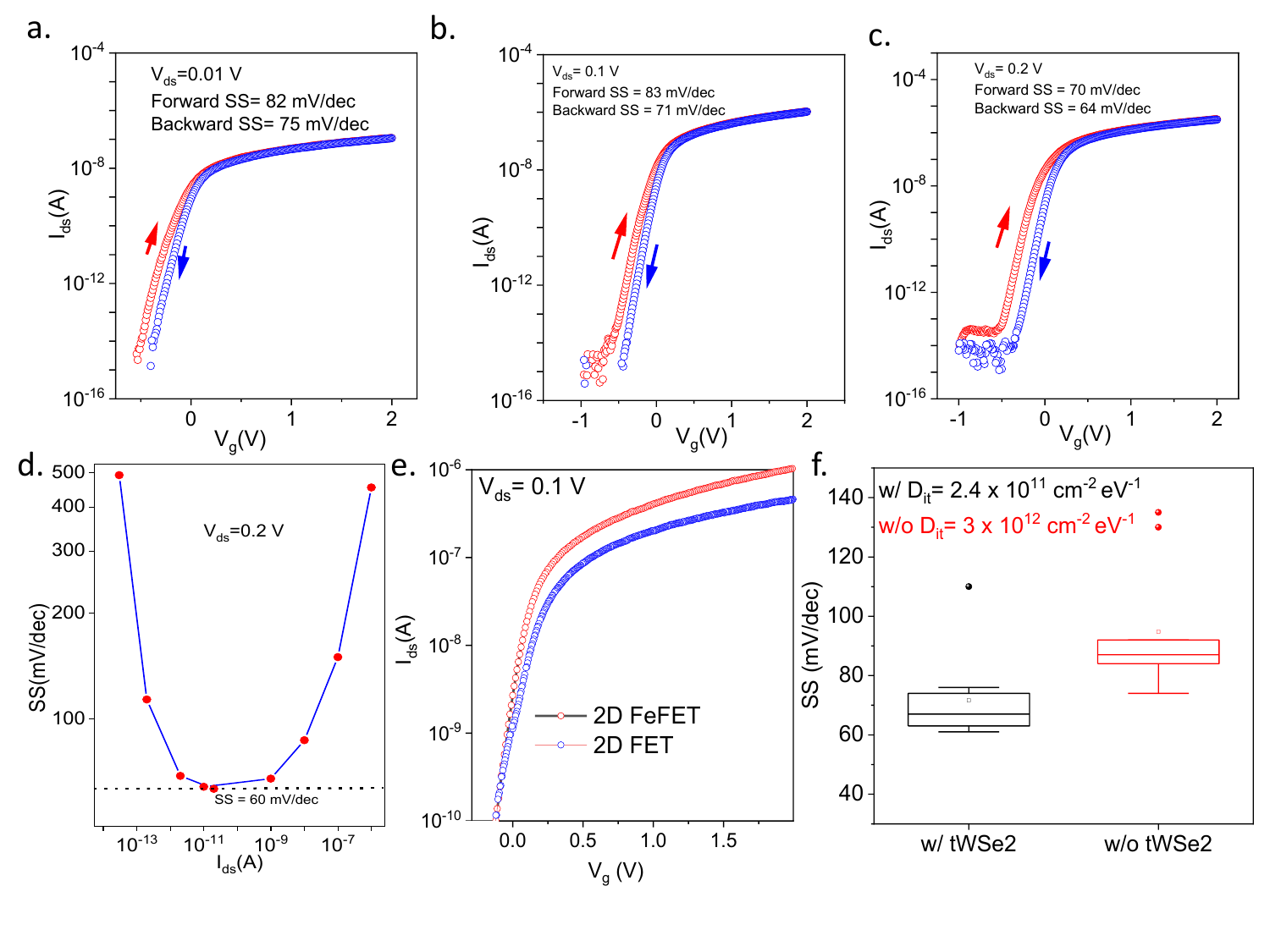} 
\caption{\textbf{Drain-bias dependence of subthreshold characteristics in 2D FETs}. (a–c) Transfer curves at increasing $V_{\mathrm{DS}}$ ($L_{\mathrm{ch}} = 1~\mu$m) showing stable hysteresis and progressive improvement in subthreshold swing (SS). (d) Point-by-point SS versus drain current ($I_{\mathrm{DS}}$), demonstrating approach to the near-ideal 60 mV dec$^{-1}$ limit across a broad $I_{\mathrm{DS}}$ range. (e) Comparison with a MoS$_2$/hBN control device, highlighting the superior SS ($\sim$64 mV dec$^{-1}$) in the FeFET versus $>$130 mV dec$^{-1}$ in the control.}
\label{fig:transport2}
\end{figure*}

\begin{figure}
\includegraphics[height=15cm]{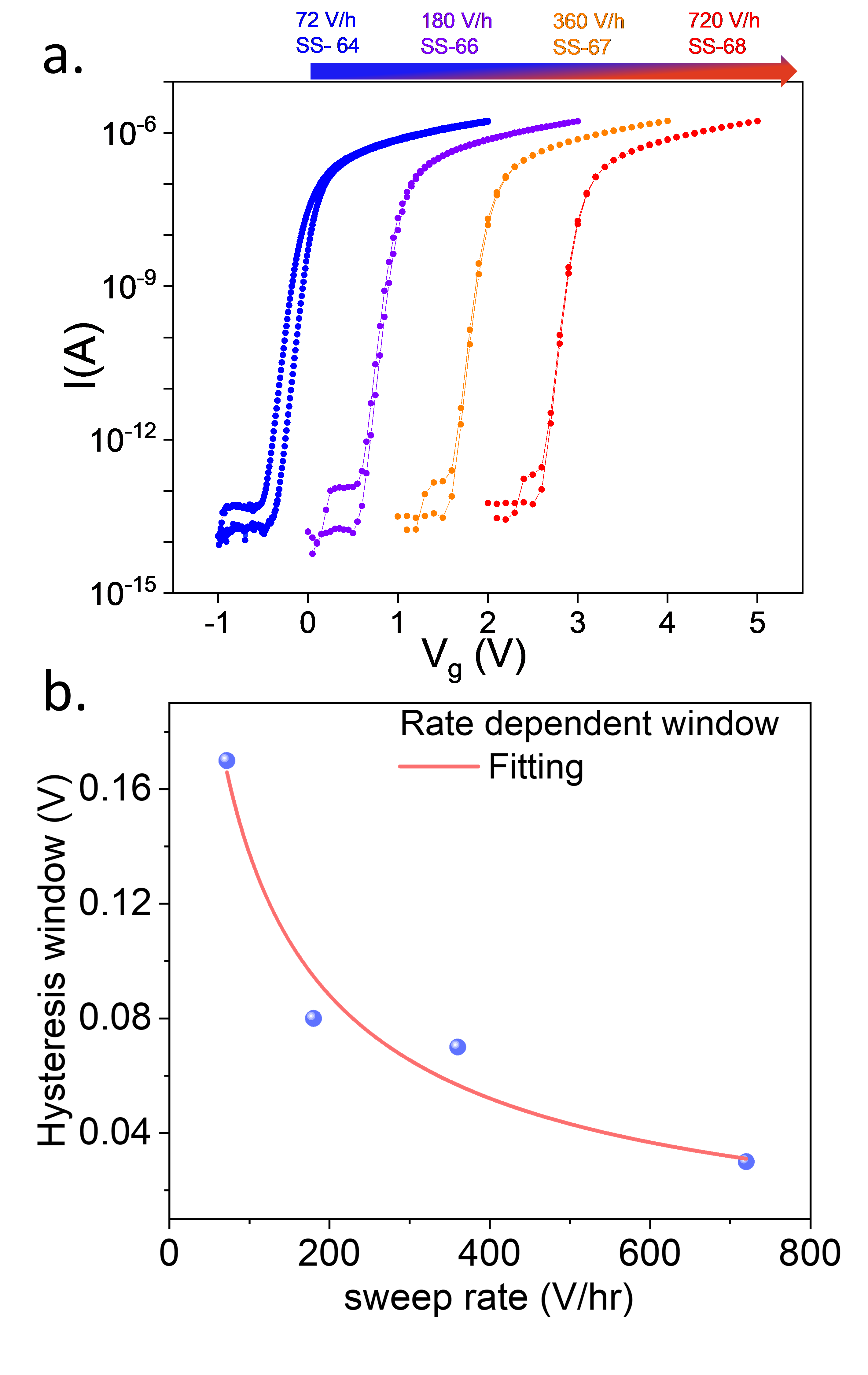}
\caption{\textbf{Sweep-rate dependence of hysteresis dynamics.} (a) Transfer curves ($I_{\mathrm{D}}$–$V_{\mathrm{G}}$) measured at different sweep rates, shifted by 1~V for clarity, showing systematic reduction of hysteresis with increasing rate. (b) Extracted hysteresis width versus sweep rate, fitted using a domain-wall-limited switching model. }
\label{fig:transport3}
\end{figure}

\begin{figure*}
\includegraphics[width=1\linewidth]{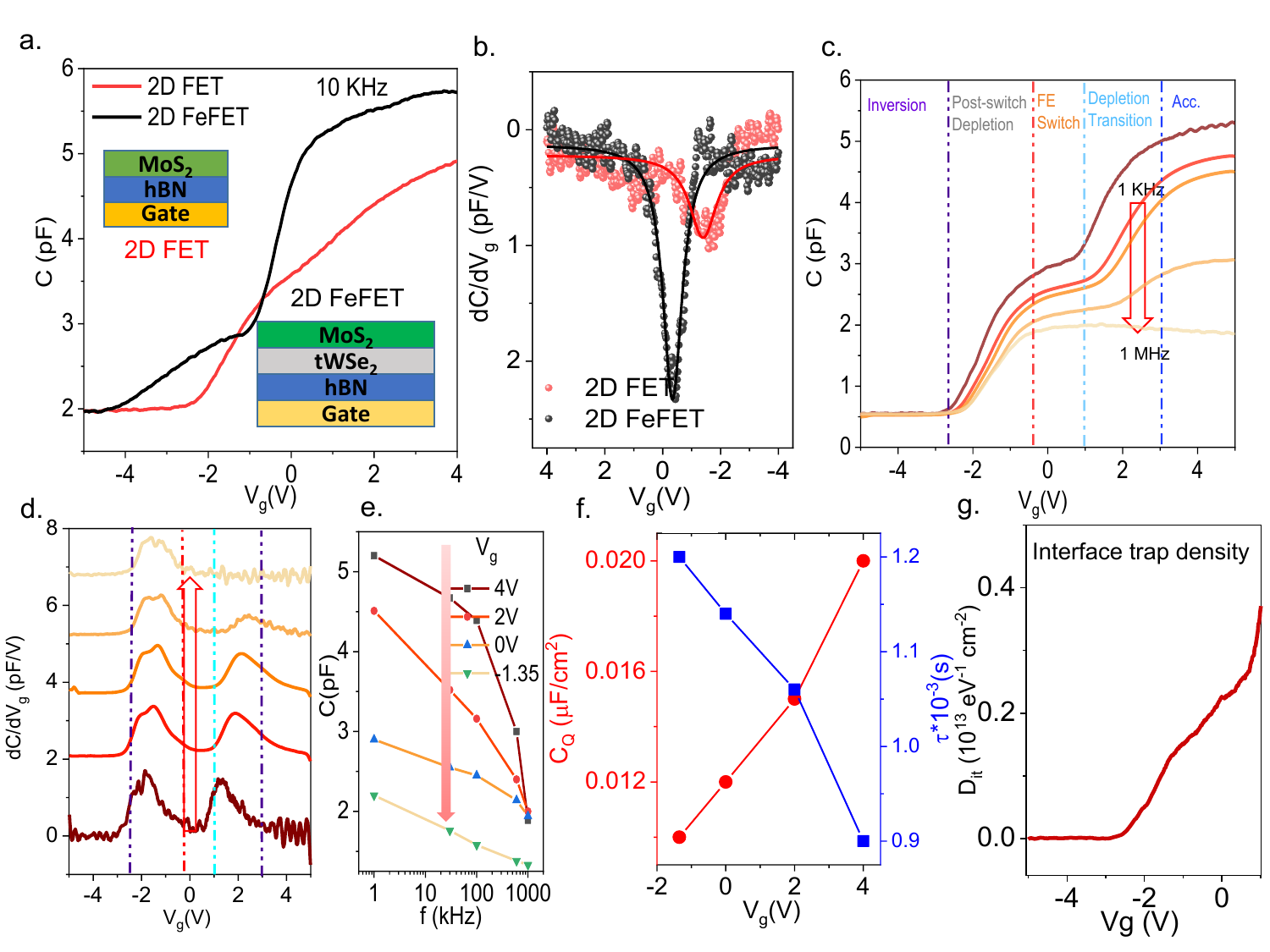} 
\caption{\textbf{CV spectroscopy of 2D FeFETs.} (a) CV characteristics of the MoS$_2$/hBN/twisted-WSe$_2$ FeFET compared with a control device, showing enhanced capacitance (inset: device schematic). (b) Differential $dC/dV_{\mathrm{G}}$ plot displaying a sharp peak for the twisted-WSe$_2$ device, indicative of ferroelectric switching, whereas the control shows a broad peak corresponding to deep depletion or trap response; Lorentzian fits are used to extract the FWHM. (c) Frequency-dependent CV curves of another FeFET reveal an additional ferroelectric switching regime beyond the conventional CV response. (d) $dC/dV_{\mathrm{G}}$ as a function of frequency showing kinetic limitation in domain wall motion. (e) Experimental $C_{\text{total}}$ as a function of frequency at different $V_{\text{BG}}$ values. (f) Extracted $\tau_{\text{it}}$ and $C_{Q}$ as a function of $V_{\text{G}}$. (g) Experimental and simulated two-dimensional CV spectroscopy maps, highlighting the agreement between measurement and self-consistent modeling.}
\label{fig:transport4}
\end{figure*}

\begin{figure*}
\includegraphics[width=1\linewidth]{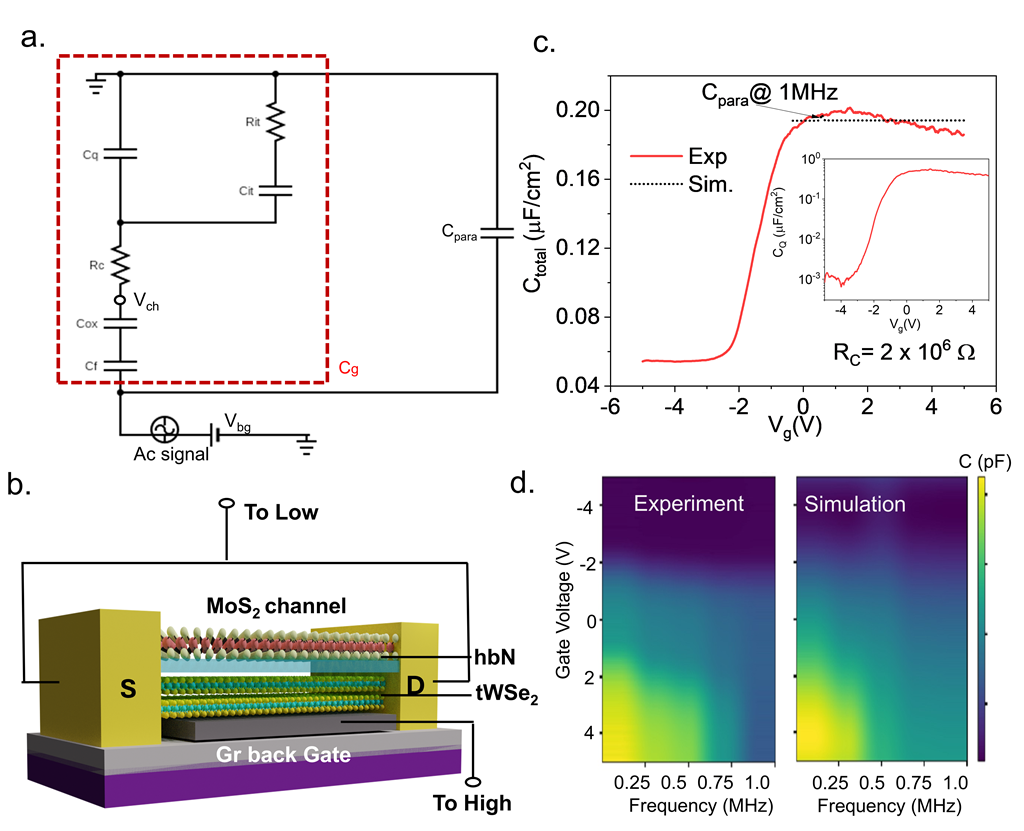}
\caption{\textbf{Simplified equivalent circuits to model MoS$_2$-FeFET C–V. } (a) Full equivalent circuit illustrating the formation of the total measured capacitance $C_{\mathrm{para}}$. 
    (b) Schematic of the capacitance measurement configuration, where both the source and drain electrodes are connected to the low terminal, and the top gate is connected to the high terminal. 
    (c) Experimental and simulated total capacitance--voltage ($C_{\mathrm{total}}$--$V_{\mathrm{TG}}$) characteristics of the MoS$_2$ FeFET, showing good agreement in the accumulation regime. 
    (d) Experimental and simulated two-dimensional $C$--$V$ spectroscopy maps, highlighting the consistency between measurement and self-consistent modeling. }
\label{fig:transport5}
\end{figure*}

\begin{figure*}
\includegraphics[width=1\linewidth]{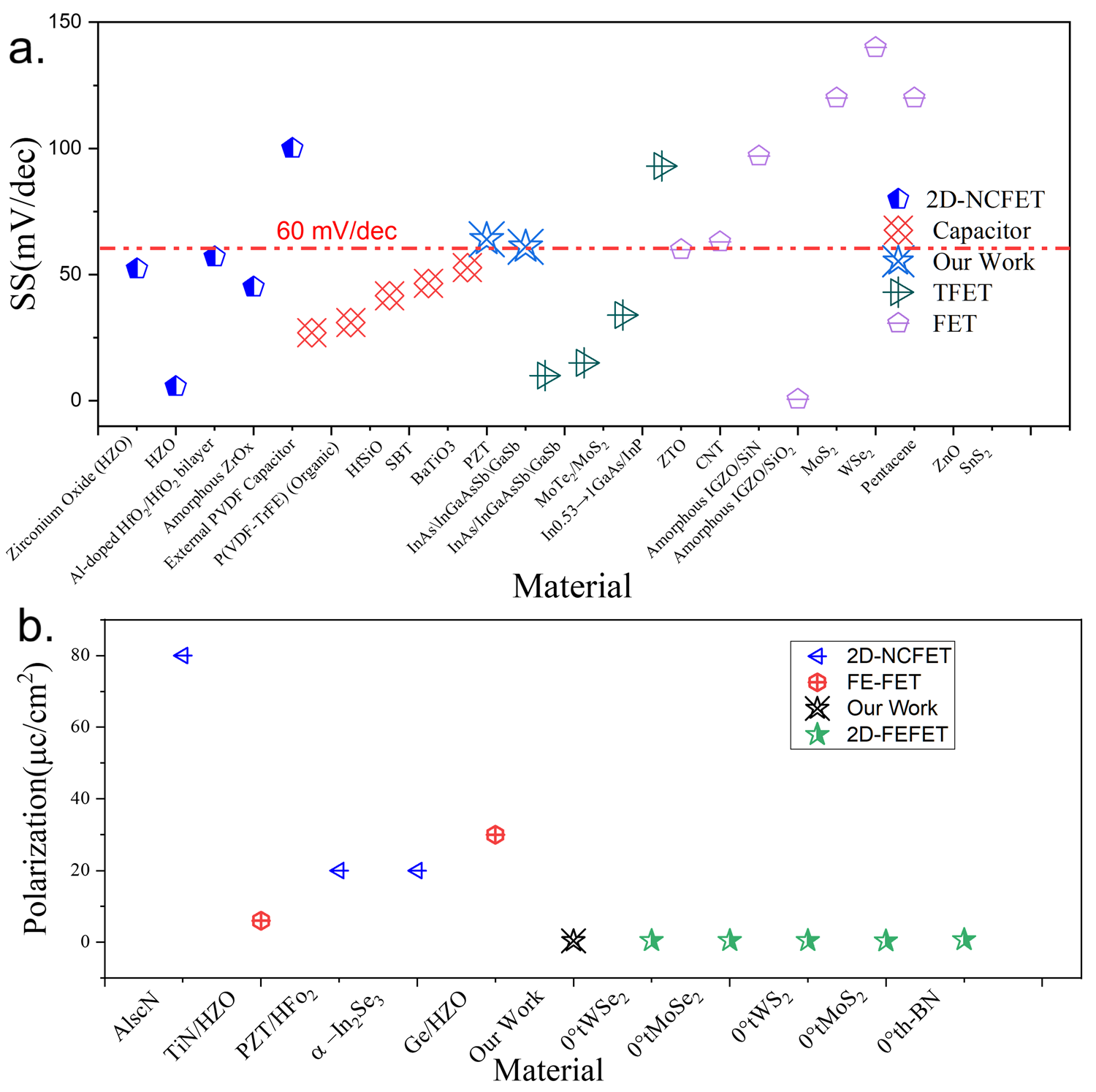} 
\caption{\textbf{Benchmarking subthreshold swing in Moiré-engineered ferroelectric transistors.} (a) Comparison of the subthreshold swing (SS) of our twisted-WSe$_2$ FeFETs with reported 2D FETs and 2D FeFETs\cite{newsom202259,song2014algan,rasool2019analytical,franklin2013consistently,zhou2012ingaas,patoary2023improvements,zhang2009low,xu2012low,nourbakhsh2016mos2,saeidi2020nanowire,cheung201060,wei2019performance,chen2014self,liu2021steep,si2018steep,lin2019subthreshold,nourbakhsh2017subthreshold,wang2018vertical,koo2018vertical,zhang2021zro}, highlighting near-ideal SS enabled by moiré-induced ferroelectric coupling and efficient polarization switching. (b) The twisted--WSe$_2$ device exhibits a remanent polarization of $P_{\mathrm{r}}\approx0.38~\mu\mathrm{C/cm^2}$ and a coercive field of $E_{\mathrm{c}}\approx0.1~\mathrm{V/nm}$, comparable to or exceeding values reported for other twisted or moiré ferroelectric systems, confirming efficient interlayer coupling and robust polarization\cite{si2019ferroelectric,yasuda2021stacking,wang2022interfacial,chung2018first,liu2021post,dasgupta2015sub,mcguire2017sustained}.}
\label{benchmark}
\end{figure*}
\newpage

\bibliography{References}
\bibliographystyle{abbrv}

\end{document}